# Interfacial Perpendicular Magnetic Anisotropy of Ultrathin Fe(001) Film Grown on CoO(001) Surface

Tong Wu[1], Yunzhuo Wu[1], Haoran Chen[1], Hongyue Xu[1], Zhen Cheng[1], Yuanfei Fan[1], Nan Jiang[1], Wentao Qin[1], Yongwei Cui[1], Yuqiang Gao[2], Guanhua Zhang[3], Zhe Yuan[4,5], Yizheng Wu[1,6,7]*

[1] *Department of Physics and State Key Laboratory of Surface Physics, Fudan University, Shanghai 200433, China*
[2] *School of Physics and Electronic Information, Anhui Normal University, Wuhu 241000, China*
[3] *State Key Laboratory of Molecular Reaction Dynamics, Dalian Institute of Chemical Physics, Chinese Academy of Sciences, Dalian, Liaoning, 116023, China*
[4] *Institute for Nanoelectronic Devices and Quantum Computing, Fudan University, Shanghai 200433, China*
[5] *Interdisciplinary Center for Theoretical Physics and Information Sciences, Fudan University, Shanghai 200433, China*
[6] *Shanghai Research Center for Quantum Sciences, Shanghai 201315, China*
[7] *Shanghai Key Laboratory of Metasurfaces for Light Manipulation, Fudan University, Shanghai 200433, China*

## Abstract

Exploring novel systems with perpendicular magnetic anisotropy (PMA) is vital for advancing memory devices. In this study, we report an intriguing PMA system involving an ultrathin Fe layer on an antiferromagnetic (AFM) CoO(001) surface. The measured perpendicular anisotropy field is inversely proportional to the Fe thickness, indicating an interfacial origin of PMA. Temperature-dependent measurements reveal that the antiferromagnetism of CoO has a negligible effect on the PMA. By leveraging the magneto-optical Kerr effect and birefringence effect, we achieved concurrent visualization of ferromagnetic (FM) and AFM domains. A pronounced coupling effect between these domains was observed near the spin reorientation transition, contrasting sharply with areas of stronger PMA that exhibited weak coupling. This research not only establishes a new FM/AFM bilayer PMA system but also significantly advances the understanding of FM/AFM interfacial interactions.

\* wuyizheng@fudan.edu.cn

# I. Introduction

Perpendicular magnetic anisotropy (PMA) is pivotal for advancing spintronic devices, offering enhanced thermal stability and reduced power consumption, which are essential for applications such as magnetoresistive random access memories [1,2]. Thus, PMA has been extensively explored in various systems, including bulk materials exhibiting strong uniaxial magnetocrystalline anisotropy [3,4], and thin films with interfacial-origin anisotropy, such as cobalt-based metallic multilayers [5–7] and ordered alloys [8]. Interfacial perpendicular magnetic anisotropy (iPMA), with its easily tunable characteristics, enables more versatile applications in devices.

Originating from the spin-orbit interaction-induced hybridization of relevant electronic orbitals, iPMA has also been observed in ferromagnet/oxide systems over the past few decades [9–13]. The discovery of iPMA in CoFeB/MgO [9,10] and Co/oxide [11] suggests that these materials are promising candidates for ferromagnetic (FM) electrodes in next-generation high-density and non-volatile memory devices. Additionally, *ab initio* calculations have predicted that the iPMA of Fe/MgO could surpass that of CoFeB/MgO, depending on the interface state [14,15]. Subsequent experiments have confirmed the presence of PMA in the Fe/MgO system [12,13], demonstrating that the enhanced anisotropic orbital angular momentum of Fe at the interface contributes to PMA [16]. Recently, the PMA was reported in a Fe/NiO bilayer grown on the MgO(001) substrate [17], with antiferromagnetic (AFM) spins of NiO aligned perpendicular to the film [18,19]. The similar rock-salt structure of NiO compared to MgO provides the opportunity for realizing orbital hybridization at the interface, contributing to the PMA. However, the relationship between the PMA of Fe and the antiferromagnetism of the adjacent layer, as well as the coupling between FM and AFM domains in such systems, remains unexplored. This is crucial because the magnetic domains serve as information carriers, and effective manipulation of memory units requires a deeper understanding of the characteristics of the system. Furthermore, in order to broaden PMA systems and explore more possibilities for designing spintronic devices, further investigation is needed to determine whether PMA is prevalent in other Fe/AFM oxide systems.

CoO is also a rock-salt antiferromagnet with a Néel temperature ($T_N$) of 293 K in its bulk phase [20]. Because the lattice constant of CoO is slightly larger than that of MgO ($a_{CoO}$ = 4.26 Å > $a_{MgO}$ = 4.21 Å), the epitaxial CoO(001) thin film on MgO(001) substrate undergoes in-plane compressive strain. This strain induces an in-plane biaxial spin alignment along the [110] and [$\bar{1}$10] directions. Additionally, it results in a higher $T_N$ of ~310 K due to strong magneto-elastic coupling [21,22]. The presence of two available in-plane states for AFM CoO domains, combined with the relatively low $T_N$, makes it a suitable system for developing energy-efficient information storage devices operable under ambient conditions.

In this paper, we demonstrate the PMA in ultrathin Fe layers deposited on the CoO(001) surface. Through quantitative measurements with varying Fe thicknesses, we confirmed the interfacial origin of PMA and identified the spin reorientation transition (SRT) thickness as approximately 0.98 nm. Notably, the determined iPMA energy is 1.38 ± 0.01 mJ/m² at 340 K (above $T_N$) and 300 K (below $T_N$), independent of the antiferromagnetism of CoO. We also explored the coupling between FM Fe domains and underlying AFM CoO domains utilizing the polar magneto-optical Kerr effect (MOKE) and the magneto-optical birefringence effect (MOB). It was revealed that FM domains are strongly coupled with AFM domains near the SRT region. However, no significant coupling between FM and AFM domains was observed in the area with stronger PMA in thinner Fe films, as also confirmed by X-ray photoemission electron microscopy (PEEM) measurements. Our experiments establish a novel PMA system based on a ferromagnet/antiferromagnet heterostructure, which provides a promising platform for future research into fascinating phenomena and spintronic devices.

## II. Experiments

Single-crystalline Fe(001)/CoO(001) films were prepared at room temperature (RT) by molecular beam epitaxy in an ultra-high vacuum system with a base pressure of 2×10⁻¹⁰ Torr [23]. The 8-nm-thick CoO layer was grown on the pre-annealed single-crystal MgO(001) substrate by reactive deposition of Co at an oxygen pressure of $4\times10^{-6}$ Torr [24,25]. The Fe layer was grown into a wedge shape on the CoO film for the thickness-dependent measurements. The epitaxial growth of the CoO and Fe layers is evident from the sharp reflection high energy electron diffraction (RHEED) patterns. The RHEED patterns were observed with the incident e-beam parallel to the [100] and [110] crystal axes of the MgO substrate, as illustrated in Figs. 1(a)-(c) and Figs. 1(d)-(f), respectively. The elongated stripe shapes in the RHEED patterns confirm the smoothness of the deposited layers with the lattice relationship of Fe<110>(001)//CoO<100>(001)//MgO<100>(001) [26]. Subsequently, a 1 nm-thick MgO layer was deposited *in situ* to prevent oxidation of the Fe layer before the sample was transferred into the magnetron sputtering system. Then, an additional capping layer was deposited for further protection. For MOKE and optical microscopy measurements, 3 nm-thick $Al_2O_3$ was deposited on top as the protection layer. For transport and PEEM measurements, a 2 nm-thick Pt capping layer was used.

Figure 1(g) shows the typical X-ray diffraction pattern for a 14 nm-thick CoO film grown on the MgO(001) substrate. The CoO peak is located at 41.92°, closely matching the MgO(002) peak at 42.96°, signifying excellent epitaxial growth. The lattice constant of CoO perpendicular to the film surface was calculated as 4.31 Å, slightly exceeding the bulk value of 4.26 Å. This increase is consistent with the expected in-plane compressive lattice strain, resulting from epitaxial growth on the MgO substrate with a lattice constant of 4.21 Å. Additionally, the observable Laue oscillations on both sides of the CoO(002) peak suggest low roughness at both the top and bottom interfaces of the CoO layer.

MOKE was employed to characterize the magnetic properties of the sample with a wedge-shaped Fe layer (0–1.2 nm) in both longitudinal and polar modes at RT. The in-plane component of the incident light in longitudinal mode was aligned parallel to the wedge direction, as shown in Fig. 2(a). The magnetic hysteresis loops of the

Fe/CoO bilayer were measured through a superconducting quantum interference device (SQUID) under an in-plane magnetic field at various temperatures.

To determine the PMA quantitatively, anomalous Hall effect (AHE) measurements were conducted on a wedge-shaped Fe sample with thicknesses ranging from 0 to 3 nm. The AHE measurements were carried out in a superconducting magnet under a rotating magnetic field, using the standard lock-in technique. The sample was initially patterned into a Hall bar with multiple Hall probes along the thickness gradient using standard lithography and $Ar^+$ etching techniques. The Hall bars have a width of 50 μm and a separation of 200 μm between adjacent Hall probes [27].

The FM domains of Fe in the demagnetized state and their interplay with the underlying AFM domains of CoO were investigated by a commercial Kerr microscope from Evico Magnetics GmbH [28,29]. The FM and AFM domains in the Fe/CoO bilayer were also imaged using PEEM based on X-ray magnetic circular dichroism (XMCD) and x-ray magnetic linear dichroism (XMLD), respectively. The XMCD-PEEM and XMLD-PEEM measurements were performed at BL09U (Dreamline) of the Shanghai Synchrotron Radiation Facility (SSRF).

## III. Results and discussion

### A. Interfacial PMA and thickness-dependent SRT in Fe/CoO(001)

Figures 2(c) and 2(d) illustrate the hysteresis loops of Fe film with different thicknesses ($d_{Fe}$) under out-of-plane (OOP) and in-plane (IP) magnetic fields, respectively. The applied IP field is along the Fe [100] direction, which is the easy axis of the Fe(001) film. The ultrathin Fe film starts to establish ferromagnetism at around 0.4 nm, below which only paramagnetic OOP and IP loops can be observed. A square-shaped OOP loop appears for $d_{Fe}$ = 0.6 nm, as shown in Fig. 2(c), exhibiting the pronounced PMA. In Fig. 2(d), the measured Kerr loop with the same thickness shows a parabolic shape with a linear background, indicating that the maximum field is insufficient to align the Fe magnetization in the film plane. In Figs. 2(c)-(d), both the OOP and the IP loops for $d_{Fe}$ = 1.0 nm display a tilted shape, suggesting weaker PMA. For $d_{Fe}$ = 1.2 nm, the OOP loop displays the typical hard-axis loop with a saturation field of 0.4 T, while the IP loop shows a square shape, confirming the IP magnetic anisotropy of the Fe film.

By extracting the remanent signal ($M_r$) at zero field and the saturation signal ($M_s$) at maximum field (0.5 T for the OOP field and 0.25 T for the IP field), the remanent ratio of $M_r/M_s$ can be determined. Fig. 2(b) shows the ratio $M_r/M_s$ as a function of $d_{Fe}$. The polar MOKE ratio stabilizes at 1 for $d_{Fe}$ < 0.9 nm, indicating strong PMA. For $d_{Fe}$ > 0.9 nm, the remanent ratio gradually decreases and approaches zero at ~ 1.2 nm, suggesting the dominance of IP magnetic anisotropy (IMA). Conversely, the remanent ratio measured by longitudinal MOKE exhibits the opposite trend, increasing from 0 to 1 with $d_{Fe}$. These observations imply a thickness-dependent SRT, with an SRT thickness $d_{SRT}$ of ~0.98 nm, as determined by the intersection point of the two datasets in Fig. 2(b). The determined $d_{SRT}$ aligns with a previous studies on MgO/Fe/MgO multilayers, where the dSRT was around 0.9 nm [13]. In both cases, the Fe films exhibit a single-crystal body-centered cubic structure. Considering the thickness of an iron monolayer (ML) is 0.14 nm, the $d_{SRT}$ determined at room temperature in our experiments corresponds to approximately 7 MLs.

In order to understand the origin of the observed PMA, we quantified the magnetic anisotropy as a function of $d_{Fe}$ by measuring the angular-dependent AHE signals based on the principle of magnetic torquemetry [30,31]. Fig. 3(a) shows the measurement geometry with the Hall bar along the Fe[100] direction. Here, we define $x$, $y$, and $z$ as the current direction, the IP direction perpendicular to the current, and the film normal direction, respectively. The magnetic field of 1.5 T was rotated in the $xz$ plane. In this geometry, the absence of planar Hall effect contribution allows for the exclusive measurement of the AHE signal. Due to the magnetic anisotropy, the 1.5 T field cannot fully align the magnetic moment along the field direction, thus the field angle $\theta_H$ and the magnetization angle $\theta_M$ could be different. In the macrospin model, the energy density $\varepsilon(\theta_M)$ of Fe film can be expressed as [31]:

$$\varepsilon(\theta_M) = -M_s H\cos(\theta_H - \theta_M) + K_2\sin^2\theta_M + K_4\sin^2\theta_M\cos^2\theta_M. \#(1)$$

The first term represents the Zeeman energy, where $M_s$ signifies the saturation magnetization of Fe. The second and third terms represent the uniaxial and four-fold anisotropy energy densities, with $K_2$ and $K_4$ as the corresponding energy constants, respectively. The equilibrium angle $\theta_M$ is obtained by minimizing the free energy with respect to $\theta_M$, and the anisotropy torque field $\tau(\theta_M)$ can be expressed as follows:
$$\tau(\theta_M) \equiv H\sin(\theta_H - \theta_M) = H_{k2}\sin 2\theta_M + H_{k4}\sin 4\theta_M, \#(2)$$
where the anisotropy fields $H_{k2}$ and $H_{k4}$ are defined as $H_{k2} = K_2/M_s$ and $H_{k4} = K_4/(2M_s)$, respectively. The relationship of $\theta_M(\theta_H)$ can be determined by the measured $R_{xy}(\theta_H)$, since $R_{xy}(\theta_H)$ is expected to be proportional to $\cos(\theta_M)$. Then, by fitting the $\tau(\theta_M)$ curve with Eq. (2), $H_{k2}$ and $H_{k4}$ can be determined.

Figure 3(b) illustrates the typical $R_{xy}(\theta_H)$ curves for different $d_{Fe}$ at 300 K, with minimum and maximum AHE signals at 0 and 180 degrees, indicating that the 1.5 T field is strong enough to align the magnetization along the z-axis. Fig. 3(c) presents the calculated $\tau(\theta_M)$ curves. The curves for Fe films with $d_{Fe}$ = 0.55 nm and 3 nm exhibit two-fold symmetry but with opposite signs, attributed to their OOP and IP anisotropy, respectively. For the Fe film with $d_{Fe}$ = 0.95 nm, which approaches $d_{SRT}$, the determined $\tau(\theta_M)$ has a very small amplitude with the superposition of two-fold and four-fold symmetry. The red line in Fig. 3(f) shows the fitted thickness-dependent $H_{k2}$ at 300 K, which is inversely proportional to $d_{Fe}$ for $d_{Fe}$ > 0.85 nm. Note that the positive $H_{k2}$ indicates the easy axis perpendicular to the film, and the sign reversal at 0.92 nm signifies the SRT from OOP to IP direction.

To investigate the effect of the antiferromagnetism of CoO on the PMA of Fe, we also conducted angular-dependent AHE measurements above the $T_N$ of CoO. The $T_N$ was determined through the temperature-dependent exchange coupling between CoO and Fe. Fig. 3(d) displays the hysteresis loops at different temperatures measured by SQUID for an Fe(5 nm)/CoO(8 nm) bilayer, with the IP magnetic field along the easy axis of the Fe film. As the temperature decreases, the exchange coupling between CoO and Fe strengthens [32], leading to an increased coercive field ($H_c$) in Fe. Fig.

3(e) shows the temperature-dependent $H_c$, which clearly decreases with temperature. It is clear that the $H_c$ shows very weak temperature-dependence for $T > 315$ K, indicating a $T_N$ of ~ 315 K, which is close to the reported value of ~ 310 K in previous work [22].

Fig. 3(f) also shows the $d_{Fe}$-dependent $H_{k2}$ measured at 340 K (above $T_N$), which is nearly the same as that measured at 300 K, indicating that the antiferromagnetism of CoO has a negligible influence on the PMA of the upper Fe layer. The $H_{k2}$ can be expressed as $H_{k2} = \frac{2K_b}{M} + \frac{2K_i}{M}d_{Fe}^{-1}$, where $K_b$ and $K_i$ represent the bulk anisotropy and interface anisotropy, respectively. The determined $H_{k2}$ at both 300 K and 340 K for $d_{Fe} > 0.85$ nm can be well fitted by the linear relationship described above. Combined with the results of thickness-dependent SRT, our results imply that the PMA in Fe/CoO(001) originates from the interface. Using the measured saturation magnetization $M$ by SQUID, the interface anisotropy $K_i$ at each temperature can be obtained. The $K_i$ value is 1.38±0.01 mJ/m² at 300 K and 1.37±0.01 mJ/m² at 340 K, respectively. Thus, the antiferromagnetism of CoO has a negligible impact on the PMA of Fe film. It is understandable that the exchange coupling field from CoO acting on Fe is isotropic within the plane perpendicular to the CoO spins and does not generate an out-of-plane uniaxial anisotropy field. As a result, the antiferromagnetic arrangement of CoO does not influence the PMA of Fe. Note that the iPMA energy at the Fe/MgO interface exhibits an upper limit of 0.38 mJ/m² [13], the iPMA energy of Fe/CoO interface is estimated to be 1.00 mJ/m², slightly larger than the value in Fe/NiO bilayer [17]. The bulk anisotropy $K_b$ is determined to be -1.52±0.01 MJ/m³ at both 300 K and 340 K.

We also performed the thickness-dependent AHE measurements at 5 K, and the determined $H_{k2}$ in Fig. 3(f) also demonstrated a good $1/d_{Fe}$ dependence for $d_{Fe} > 0.6$ nm at low temperature. Considering the presence of charge transfer at the interface in the MgO/Fe/MgO system [33], which has a similar structure to our sample, charge transfer is also expected to occur in our system. Charge transfer can influence the PMA by altering the strength of the spin-orbit coupling at the interface. Therefore, based on the linear relationship between $H_k$ and $1/d_{Fe}$ for $d_{Fe}$ exceeding 0.6 nm, it can be inferred that charge transfer at the interface in our system becomes saturated when $d_{Fe}$ exceeds 0.6 nm. The linear fitting yields the values of $K_i$ and $K_b$ to be $1.98 \pm 0.01$ mJ/m$^2$ and $-1.78 \pm 0.01$ MJ/m$^3$, respectively, demonstrating the enhancement of magnetic anisotropy at low temperature. Note that the effective anisotropy energy $K_u$ can be written as $K_u = \frac{K_i}{d_{Fe}} + K_b$, where $K_b$ is dominated by the demagnetization energy. The calculated demagnetization energy of -1.81 MJ/m$^3$ aligns well with the determined bulk-limit anisotropy energy $K_u$ ($K_u = K_b$), confirming that the iPMA contribution vanishes in the bulk limit. It should also be noted that $H_{k2}$ changes its sign at $d_{SRT} \sim 1.02$ nm, which is thicker than the $d_{SRT}$ of 0.92 nm at RT. Therefore, for the Fe films with the thicknesses between 0.92 nm and 1.02 nm, the temperature-driven SRT could be observed.

**B. Exchange coupling between FM and AFM domain in Fe/CoO(001)**

Exchange coupling is essential in the applications of magnetic sensors and magnetic memories [34]. Previous studies have extensively investigated the exchange coupling in AFM/FM bilayers, where both layers exhibit PMA or IMA [32,34–38]. However, domain coupling between a ferromagnet with PMA and an antiferromagnet with IMA has been rarely studied, especially in cases where the ferromagnet undergoes thickness-dependent SRT.

We first employed polar-MOKE microscopy to image the evolution of FM domains as a function of $d_{Fe}$. Figures 4(a)-(h) show the demagnetized states of FM domains for increasing $d_{Fe}$, realized by applying an oscillating IP magnetic field decaying from 0.7 T. To minimize the influence of CoO domains on FM domain imaging, the oscillating IP field was aligned along the CoO[110] direction, so that the reversible switching of FM domains will not change the CoO domain configuration [29]. As a result, the CoO layer forms an IP single AFM domain with the Néel vector (***n***) along the [$\bar{1}$10] direction due to the spin-flop coupling with the Fe spins [29,38]. The polarization direction of the incident light, indicated by the blue double arrow in

Fig. 4(a), is parallel to the CoO[110] direction. Consequently, a zero MOB effect is expected [39,40], and the measured domain contrast in Fig. 4 arises solely from the polar magnetic signal of the Fe film. Determined from the field-dependent measurement, the bright and dark contrasts are identified as representing the magnetizations oriented in the +z and -z directions, respectively.

The FM domains are distinctly observable at a reduced $d_{Fe}$ of 0.44 nm, illustrated in Fig. 4(a). The domains exhibit irregular shapes with widths of approximately 1 μm, with their positions corresponding to nucleation sites during the demagnetization process. The high density of nucleation sites can be attributed to the weak PMA and the inherent surface roughness of ultra-thin films. Fig. 4(b) depicts a reduction in nucleation sites with an enlargement of domain size, yet irregular domain walls due to robust pinning persist. Figs. 4(c)-(d) show larger domains within thicker Fe films, indicating a substantial reduction in pinning effects caused by defects. As $d_{Fe}$ increases further, labyrinthine domains emerge, as shown in Figs. 4(e)-(g). The domain size decreases while the density increases, consistent with the weakened PMA. Fig. 4(h) presents domains near the SRT region at $d_{Fe}$ = 0.98 nm. The single-domain state on the right side of the image signifies the IP magnetic anisotropy, whereas the PMA region on the left exhibits numerous elliptical, bubble-like, and ring-like domains. The inset highlights a ring-like domain, marked by the orange box, resembling a skyrmion bag as described in Refs. [41,42]. The reported interfacial Dzyaloshinskii-Moriya interaction (iDMI) at the FM-oxide interface [43,44] may also be present in our system. Thus, the bubble-like and ring-like domains near the SRT region are likely topologically-protected spin textures, necessitating additional research in the future.

We next examined the exchange coupling between FM and AFM domains in regions with strong PMA ($d_{Fe}$ = 0.80 nm) and near the SRT ($d_{Fe}$ = 0.98 nm), employing both the polar-MOKE and the MOB effect. The MOB effect has been confirmed for its ability to image the AFM domains and their switching dynamics under external fields [25,29,39,40,45]. For both polar-MOKE and MOB measurements, linearly polarized light is incident perpendicular to the sample surface. While the MOKE signal is independent of the polarization direction of the incident light, the MOB signal varies with *sin2φ*. Here, *φ* denotes the angle between the polarization direction and ***n*** [39]. By rotating the polarization of the incident light, we can separate the signals from FM and AFM domains. As depicted in Fig. 5(a), varying polarization directions results in the same Kerr rotation angle ($+\theta_k$) for the reflected light, producing uniform contrast on the CCD camera. In the MOB measurement as illustrated in Fig. 5(b), the linearly polarized light is oriented at ±45 degrees relative to ***n***. Accordingly, the birefringence angles of the reflected light, while equal in magnitude, exhibit opposite signs, resulting in distinct contrasts. By adding and subtracting images obtained with light polarized at ±45 degrees relative to ***n***, we are able to separate pure FM and AFM domains.

In regions with strong PMA, an initial magnetic field was applied along the +z direction to establish a single FM domain state. Subsequently, an IP oscillating decaying magnetic field ($H_{dem}$) of less than 1000 Oe, was applied along the CoO[110] direction to achieve a partially demagnetized state, as shown in Fig. 5(e). The left side in Fig. 5(e) maintains a single-domain state due to the strong PMA, remaining unaffected by the IP magnetic field. The incident light was polarized at an angle of +45 degrees relative to the +x direction, as indicated in the lower left corner of Fig. 5(e). The oscillating field $H_{dem}$ along the [110] direction aligned the Néel vector of CoO in the [$\bar{1}$10] direction, forming a single domain with uniform contrast. Next, the IP external field ($H_{ext}$) was gradually increased along the [1$\bar{1}$0] direction until part of the AFM spins rotated by 90 degrees, as indicated by the orange dashed line in Fig. 5(f). The IP field induced minor changes in the labyrinthine domains in Fe film, but no significant coupling was observed between the CoO and Fe domains. The CoO domain walls traversed the Fe multi-domain regions without being affected. To further assess the coupling between OOP FM and IP AFM domains, we extracted and compared their shapes. We then rotated the polarization direction of the incident light by 90 degrees, as depicted in Fig. 5(g). The signal from CoO inverts, while the Fe signal remains unchanged, following the principle described in Figs. 5(a)-(b). The CoO AFM domain in Fig. 5(h) was obtained by subtracting Fig. 5(f) from Fig. 5(g), and the pure Fe FM domain shown in Fig. 5(i) was determined by adding Fig. 5(f) to Fig. 5(g). The distinct patterns of AFM and FM domains indicate a weak interaction between the IP CoO domains and Fe domains with strong PMA, as illustrated schematically in Fig. 5(c).

Similar measurements were performed near the SRT region. Fig. 5(j) shows demagnetized FM domains superimposed on a single AFM domain. Figs. 5(k) and 5(l) depict the partially switched FM and AFM domains, imaged with the light polarized at +45º and -45º, respectively. Fig. 5(m) shows the calculated CoO AFM domains, and Fig. 5(n) shows the separated contrast from Fe FM domains. The switching regions of the AFM and FM domains exhibit similar shapes, highlighted by the orange dashed lines in both figures. This suggests a significant coupling between FM and AFM domains near the SRT region compared to regions with stronger PMA. Previous work has demonstrated that FM and AFM domains in IP anisotropic the Fe/CoO bilayer are strictly coupled and one-to-one correlated [29]. Accordingly, we conclude that variations in the IP components of Fe magnetization influence the coupling strength between FM and AFM domains. Larger IP components lead to enhanced domain coupling. Notably, three distinct contrast levels are observed in Fig. 5(n). The switched Fe domains (blue circle) appear slightly darker than the original brighter areas (red circle). This subtle contrast variation likely arises from different directions of the IP components of Fe. The Kerr microscope from Evico Magnetics GmbH, operating in polar mode, employs normally incident light synthesized from two oblique beams [46]. This synthesis is expected to produce a wave vector perfectly perpendicular to the film surface. However, in practice, any intensity imbalance between the beams introduces a net IP component in the wave vector, enabling detection of the IP magnetization component. Near the SRT region, the FM spins of Fe exhibit a larger IP component that couples with CoO spins. Therefore, when $H_{ext}$ along [1 $\bar{1}$ 0] was applied to rotate the ***n*** of CoO to the [110] direction, the IP magnetization component of Fe also rotated from its initial [110] (or [$\bar{1}\bar{1}$0]) direction to the [1$\bar{1}$0] direction, as illustrated in Fig. 5(d). The variation in the IP magnetization component of Fe results in a slight difference in contrast. In the PMA region shown in Fig. 5(i), the FM magnetization is perpendicular to the sample surface, with no IP components. As a result, only two types of contrast are observed.

The coupling between FM and AFM domains was further examined using XMCD-PEEM and XMLD-PEEM measurements with different $d_{Fe}$. The X-ray was incident at a 16º angle relative to the sample surface, with the IP projection aligned along the easy axis of the Fe film. For XMCD-PEEM measurements, left circularly polarized light was used. For XMLD-PEEM measurements, linearly polarized light with its polarization direction parallel to the sample surface was employed, as indicated by the purple arrows in Fig. 6. The FM domains were imaged using energy-dependent XMCD-PEEM at the $L_3$ and $L_2$ edges of Fe and Co. It has been reported that the Fe film metallizes the interfacial Co ions at the Fe/CoO interface, inducing interfacial Co FM spins, thereby accounting for the necessity of detecting the Co XMCD signal [47,48]. The CoO AFM domains were imaged using XMLD-PEEM at the Co $L_3$ edge with two energies, $E_1$ = 776.5 eV and $E_2$ = 777.0 eV [49]. The domain images were obtained by dividing the images captured at higher and lower energies. The ±z components of the FM magnetizations are represented by the bright and dark contrasts in XMCD images, as illustrated by the grayscale bar.

Figure 6(a) displays the stripe-like domains of Fe in the PMA region ($d_{Fe}$ = 0.6 nm), while Fig. 6(b) shows the FM domains of Co in the same area. The consistent signs of Fe and Co signals indicate FM coupling between Fe and the uncompensated Co at the Fe/CoO interface. Notably, the red dashed circle in Fig. 6(b) highlights an additional pattern not observed in Fig. 6(a). This pattern matches the CoO AFM domains in Fig. 6(c), implying that the uncompensated Co spins are coupled with both the OOP Fe spins and the IP AFM spins of CoO. Consequently, the uncompensated Co spins are expected to adopt a canted state, exhibiting both OOP and IP components. The varying orientations of CoO spins alter the IP orientations of Co spins via orthogonal coupling. Since XMCD signal is sensitive to the magnetic moment component along the X-ray direction, the different IP orientations of Co result in different XMCD signals, accounting for the additional pattern observed in Fig. 6(b). However, the CoO AFM domains exert no direct influence on the spin alignment in the Fe layer, confirming weak domain coupling between CoO and Fe in the strong PMA region. In the SRT region ($d_{Fe}$ = 0.9 nm), the pattern of CoO AFM domains is imprinted on both the Co and Fe FM domains, as shown by the areas marked with orange dashed lines in Figs. 6(d)-(f). It should be noted that four different grayscale values are observed in Fig. 6(d), similar to the highlighted area in Fig. 6(b), also demonstrating the canted state of Fe. This indicates that Fe spins with OOP components could tilt in four different IP directions, depending on the orientation of underlying CoO domains. As a result, the FM spins with different IP components projected along the X-ray direction change, resulting in different contrasts. These observations corroborate our magneto-optical measurements. The complex coupling behaviors between FM and AFM domains near the SRT region warrant further exploration in future research.

### IV. Summary

In summary, our study demonstrated that the iPMA in an ultrathin Fe layer epitaxially grown on an AFM CoO(001) surface. We quantified the iPMA energy to be 1.38 mJ/m² and identified the SRT thickness as 0.98 nm at room temperature. Notably, our findings reveal that the observed iPMA is independent of the antiferromagnetism of the CoO layer. Furthermore, we delved into the coupling behavior between FM and AFM domains in the CoO/Fe bilayer with different $d_{Fe}$. By combining MOKE and MOB effect, we concurrently imaged the FM and AFM domains of Fe and CoO, respectively. Our results demonstrate strong coupling between FM and AFM domains near the SRT region, which weakens in areas where the Fe layer exhibits significant PMA. Our work establishes a novel PMA system based on the Fe/AFM insulator heterostructure and highlights the complex domain coupling between FM and AFM layers. This provides a solid foundation for future investigations in spintronics, with potential applications in high-density, energy-efficient memory and logic devices.

**Acknowledgements:**

The work was supported by the National Key Research and Development Program of China (Grant No. 2022YFA1403300), the National Natural Science Foundation of China (Grant No. 11974079, No. 12274083, No.12434003 and No. 12221004), the Shanghai Municipal Science and Technology Major Project (Grant No. 2019SHZDZX01), and the Shanghai Municipal Science and Technology Basic Research Project (Grant No. 22JC1400200 and No. 23dz2260100). Yuqiang Gao acknowledges support by the National Natural Science Foundation of China (Grant No. 12404255).

Figures:

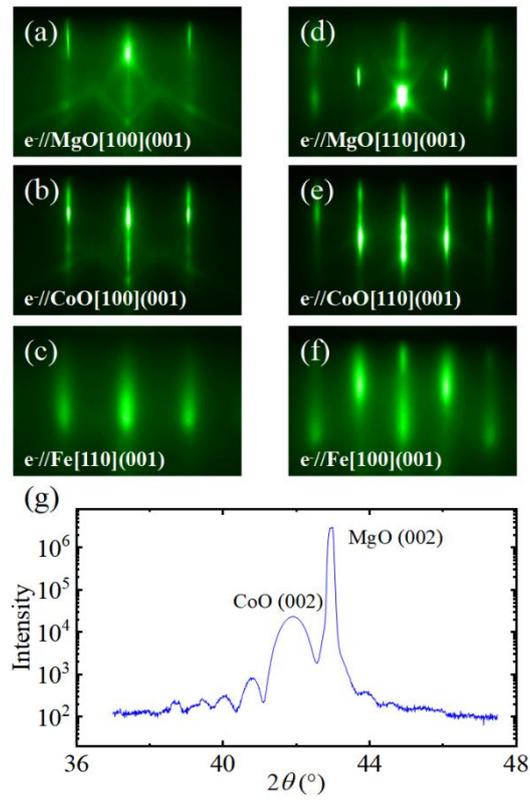

**Fig. 1.** (a)-(c) RHEED patterns of (a) MgO(001) substrate, (b) an 8 nm CoO(001) film, and (c) a 1.2 nm Fe(001) film, with the electron beam incident along the MgO[100] direction. (d)-(f) RHEED patterns of (d) the MgO(001) substrate, (e) an 8 nm CoO(001) film and (f) a 1.2 nm Fe(001) film, with the electron beam incident along the MgO[110] direction. (g) A typical XRD curve obtained from a CoO(14 nm)/MgO(001) sample.

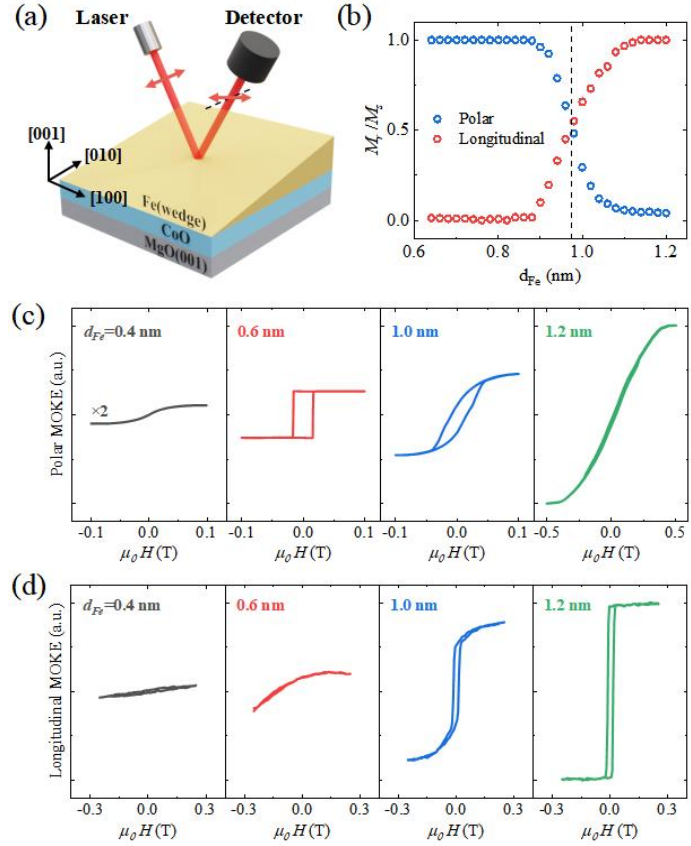

**Fig. 2.** (a) The schematic of the experimental setup for MOKE measurements on a wedge-shaped sample. The magnetic field can be oriented either in-plane or perpendicular to the film. (b) The $M_r/M_s$ ratio as a function of Fe thickness, as determined by longitudinal-MOKE and polar-MOKE techniques. (c)-(d) Typical hysteresis loops measured by (c) polar-MOKE and (d) longitudinal-MOKE for different Fe thicknesses.

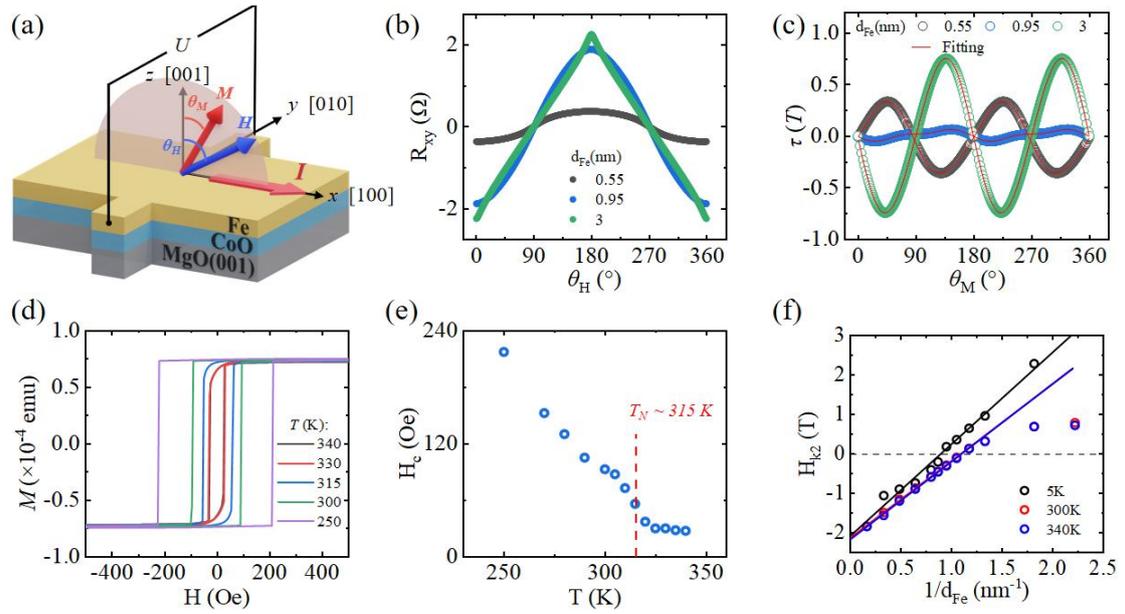

**Fig. 3.** (a) The experimental geometry for AHE measurement with rotation of the magnetic field. (b) The typical $\theta_H$-dependent AHE signals with different thicknesses of Fe. (c) The corresponding torque field curves derived from the AHE data in (b), with the red lines representing the fitting results. (d) Typical hysteresis loops of a Fe(5 nm)/CoO(8 nm) bilayer, measured by SQUID under an in-plane field at different temperatures. (e) Coercive field $H_c$ as a function of temperature, extracted from the hysteresis loops in (d). (f) Uniaxial anisotropy field $H_{k2}$ as a function of $1/d_{Fe}$ measured at different temperatures.

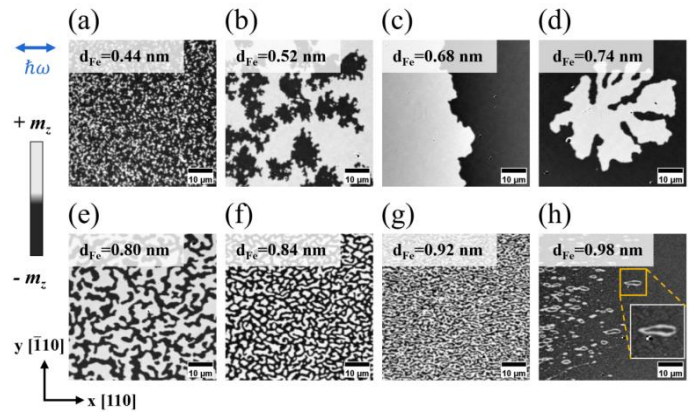

**Fig. 4.** (a)-(h) Typical Kerr domain images of Fe with different thicknesses measured in the demagnetization states. The sample was demagnetized using an oscillating in-plane field along the <110> direction, with the amplitude decaying from 0.7 T to zero. The blue double arrow in (a) indicates the polarization direction of the incident linearly polarized light. The grayscale bar represents the positive and negative components of the magnetic moment along the z-axis. The inset in (h) shows an enlarged image of the area marked by the orange square.

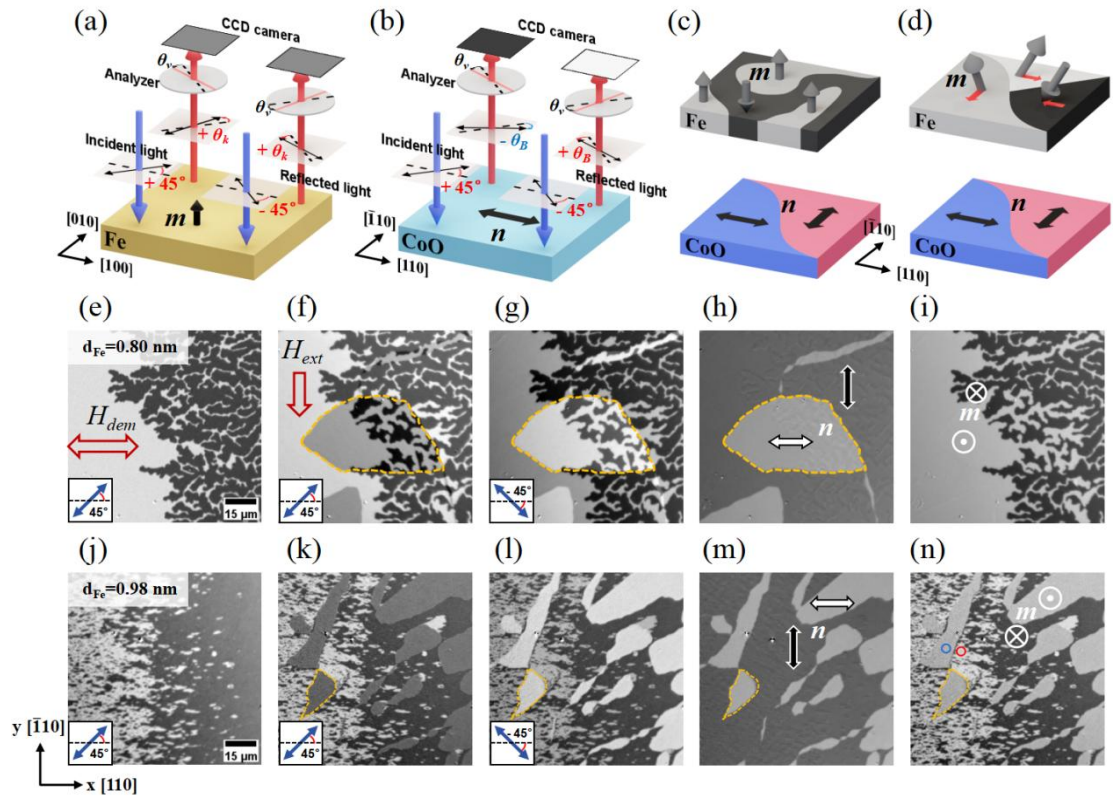

**Fig. 5.** (a)-(b) Schematics of the magneto-optical microscopy measurement geometry for (a) the Kerr effect and (b) the birefringence effect. (c)-(d) Schematic diagrams of FM and AFM domains showing (c) weak coupling in the PMA region and (d) strong coupling in the SRT region. In (d), the in-plane component of Fe spins is strongly coupled perpendicularly to the CoO AFM spins through spin-flop coupling. (e)-(n) The imaging process for separating the AFM and FM domains for Fe thicknesses of (e)-(i) 0.8 nm and (j)-(n) 0.98 nm. (e) and (j) The FM domains imaged with light polarized at +45 degrees. (f) and (k) The superimposed FM and AFM domains imaged with light polarized at +45 degrees after applying an in-plane field along the $y$-axis. (g) and (l) The domains imaged with light polarized at -45 degrees. The images shown in (h) and (m) are the derived AFM domains, while those in (i) and (n) are the derived FM domains. The orange dashed lines serve as visual guides to highlight the AFM domain boundaries and illustrate the different coupling effects for $d_{Fe}$ = 0.8 nm and $d_{Fe}$ = 0.98 nm.

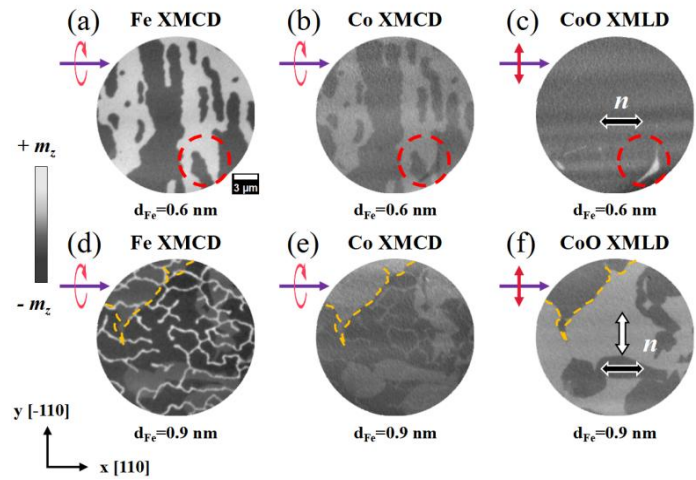

**Fig. 6.** (a)-(b) XMCD-PEEM images of FM domains for (a) Fe and (b) Co in the same area, with $d_{Fe}$= 0.6 nm. (c) XMLD-PEEM image of the AFM domains of CoO under the 0.6 nm-thick Fe layer. (d)-(e) XMCD-PEEM images of FM domains of (d) Fe and (e) Co in the same area, with $d_{Fe}$= 0.9 nm. (f) XMLD-PEEM image of the AFM domains of CoO under the 0.9 nm-thick Fe layer. The red dashed circles in (a)-(c) show the areas of CoO AFM domains, and the orange dashed lines in (d)-(f) indicate typical CoO AFM domain walls. All images are 20 μm in size.